\documentclass[twocolumn,showpacs,preprintnumbers,amsmath,amssymb]{revtex4}   
\usepackage{graphicx}% Include figure files   
\usepackage{dcolumn}% Align table columns on decimal point   
\usepackage{bm}% bold math   

\begin{document}   
%\preprint{APS/123-QED}   
   
\title{Calorimetric and transport investigations of\\ 
CePd$_{2+x}$Ge$_{2-x}$ ($x=0$ and $0.02$) up to 22~GPa}   
   
\author{H. Wilhelm}   
\affiliation{D\`{e}partement de Physique de la Mati\`{e}re Condens\'{e}e,    
Universit\'e de Gen\`eve, Quai Ernest--Ansermet 24, 1211 Geneva 4, Switzerland}   
 
\affiliation{Max--Planck--Institute for Chemical Physics of Solids,   
N\"othnitzer Str. 40, 01187 Dresden, Germany}   
   
\author{D. Jaccard}   
\affiliation{D\`{e}partement de Physique de la Mati\`{e}re Condens\'{e}e,    
Universit\'{e} de Gen\`{e}ve, Quai Ernest--Ansermet 24, 1211 Geneva 4, Switzerland}   
   
\date{23 April 2002}    
   
%################################################################   
%   
%           ABSTRACT    
%   
%################################################################   
   
\begin{abstract}   
The influence of pressure on the magnetically ordered  
CePd$_{2.02}$Ge$_{1.98}$ has been investigated by a combined  
measurement of electrical resistivity, $\rho(T)$, and ac-calorimetry,  
$C(T)$, for temperatures in the range 0.3~K~$<T< 10$~K and pressures, $p$, up to  
22~GPa. Simultaneously CePd$_2$Ge$_2$ has been examined by $\rho(T)$  
down to 40~mK. In CePd$_{2.02}$Ge$_{1.98}$ and CePd$_2$Ge$_2$ the  
magnetic order is suppressed at a critical pressure $p_c=11.0$~GPa and  
$p_c=13.8$~GPa, respectively. In the case of CePd$_{2.02}$Ge$_{1.98}$  
not only the temperature coefficient of $\rho(T)$, $A$, indicates the  
loss of magnetic order but also the ac-signal $1/V_{\rm ac}\propto  
C/T$ recorded at low temperature. The residual resistivity is  
extremely pressure sensitive and passes through a maximum and then a  
minimum in the vicinity of $p_c$. The $(T,p)$ phase diagram and the  
$A(p)$-dependence of both compounds can be qualitatively understood in  
terms of a pressure-tuned competition between magnetic order and the  
Kondo effect according to the Doniach picture. The temperature--volume
$(T,V)$ phase  
diagram of CePd$_2$Ge$_2$ combined with that of CePd$_2$Si$_2$ shows  
that in stoichiometric compounds mainly the change of interatomic  
distances influences the exchange interaction. It will be argued that  
in contrast to this the much lower $p_c$-value of  
CePd$_{2.02}$Ge$_{1.98}$ is caused by an enhanced hybridization  
between $4f$ and conduction electrons.  
\end{abstract}   
   
%high pressure, valence fluct. & HF, C(T) of mgn systems, transport, 
%h.p. apparatus) 
  
%\keywords{high pressure; ac-calorimetry; electrical resistivity; magnetic order; strong correlations}   
%Use showkeys class option if keyword display desired   
   
\maketitle    
   
%################################################################   
%   
%           INTRODUCTION   
%   
%################################################################   
   
\section{\label{sec:introduction}Introduction}   
   
The application of external pressure on metals with strong  
correlations is an established technique to tune their ground state  
properties. The electronic interactions in heavy-Fermion (HF)  
compounds can be influenced in such a way that high pressure favors  
the Kondo interaction in Ce-compounds  
\cite{Jaccard92,Movshovich96,Grosche96,Mathur98,Raymond99,Wilhelm99,Demuer00,Wilhelm00,Hegger00}  
and the RKKY interaction in Yb-systems \cite{Alami98,Knebel01}. Thus,  
pressure suppresses (favors) long range magnetic order and enhances  
(weakens) the screening of the localized 4$f$-electrons. If both  
interactions are of similar strength in the vicinity of a critical  
pressure $p_c$ often a deviation from the Fermi-liquid (FL) behavior  
is observed and some Ce-compounds even attain a superconducting ground  
state.  
   
Electrical resistivity measurements as a function of temperature, $\rho(T)$, are the standard  
method to explore the low temperature phase diagram of HF systems up  
to pressures $p\approx 20$~GPa. It is desirable to measure  
thermodynamic quantities, notably the specific heat $C(T)$ in these  
extreme conditions. The accessible pressure range for specific heat  
experiments was limited to 2-3~GPa since adiabatic techniques  
demand large sample masses and thus, a large cell volume. For an  
anvil-type of high pressure cell a much smaller sample volume is  
required, which makes an adiabatic measurement a hopeless  
venture. Among the non-adiabatic (or dynamic) methods,  
ac-calorimetry \cite{Sullivan68,Eichler79} is a suitable technique to  
be used at high pressures. Very high sensitivity can be achieved,  
whereas the absolute accuracy is less than for adiabatic  
techniques.  
   
A major step towards measuring $C(T)$ under extreme conditions has  
been achieved by implementing the ac-technique in a Bridgman-type of  
pressure cell suited for 10~GPa \cite{Bouquet00}. The sample was  
embedded in a soft mineral (steatite) and an ac-current was supplied  
to a heater close to the sample. The experimental findings have been  
confirmed by an independent study using a diamond anvil cell with He  
as pressure transmitting medium and laser heating  
\cite{Demuer00,Salce00}. Motivated by these results we assembled the  
ac-calorimetry in a Bridgman-type of high pressure cell capable of  
reaching 25~GPa and temperatures of the order of 30~mK \cite{Jaccard98}.  
   
So far, only CeCu$_2$Ge$_2$ and CeRu$_2$Ge$_2$, where Ge is an  
isoelectronic substitute for Si, have been studied extensively under  
pressure. The former compound exhibits a phase diagram similar to that  
of CeCu$_2$Si$_2$ but shifted by 9.4~GPa \cite{Jaccard99}. Close to  
the critical pressure, $p_c$, the long range magnetic order is suppressed and  
superconductivity appears like in CeCu$_2$Si$_2$ at low pressure  
\cite{Bellarbi84}. The temperature--volume $(T,V)$ phase diagram of CeRu$_2$Ge$_2$ is  
identical to that of the solid-solution CeRu$_2$(Si$_{1-x}$Ge$_x$)$_2$  
\cite{Wilhelm99}. In contrast to CeCu$_2$Ge$_2$ no superconductivity  
is observed around the magnetic/non-magnetic borderline as in  
CeRu$_2$Si$_2$ at ambient pressure. These observations support the  
notion that the Ge-substitution mainly has the effect of reducing the  
hybridization between the $4f$ and conduction electrons due to the expansion of 
the unit cell volume. This argument  
seems not to be limited to compounds crystallizing in the  
ThCr$_2$Si$_2$-type of structure. Another example is the magnetically  
ordered CeCu$_5$Au, which can be pressure-tuned into a non-magnetic  
ground state, analo\-gous to the HF prototype CeCu$_6$ at ambient  
pressure \cite{Wilhelm01}. The pressure study revealed a deviation  
from FL behavior and a low temperature anomaly in $\rho(T)$ close to  
$p_c$ which could be interpreted as faint traces of a superconducting  
state \cite{Wilhelm00}.  
   
All these studies have in common that pressure has been applied to  
stoichiometric compounds. Small deviations from stoichiometry are  
believed to result in strong effects on the electronic  
correlations. Detailed investigations of the influence of Ni-excess in  
Ce$_{1.005}$Ni$_{2+z}$Ge$_{2-z}$ have shown that a low residual  
resistivity, $\rho_0$, is a crucial requirement for the occurrence of  
incipient superconductivity at ambient pressure \cite{Steglich00} and  
that the transition temperature can be shifted upwards by carefully  
adjusting the Ni excess \cite{Jaccard01}.  For the stoichiometric  
sample, however, pressure had to be applied to achieve  
superconductivity \cite{Braithwaite00}. The combination of these  
results have led to the hypothesis that an enhanced hybridization due  
to an electronically different environment of the Ce-ions is a crucial  
ingredient to reduce $p_c$ and to achieve a superconducting ground  
state.  
   
In this article we report on results of $\rho(T)$ as well  
as ac-calorimetry experiments on CePd$_{2}$Ge$_2$ and $\rho(T)$ measurements on  
CePd$_{2.02}$Ge$_{1.98}$ performed in {\it one} pressure experiment.  
CePd$_2$Ge$_2$ enters an antiferromagnetically ordered phase at  
$T_N\approx 5.1$~K \cite{Fak00,Fukuhara98,Iwasaki87,Besnus92}. Its  
Si-counterpart, the HF system CePd$_2$Si$_2$ exhibits a similar  
magnetic structure with $T_{\rm N}\approx 10$~K  
\cite{Thompson86,Steeman88}. Several groups have confirmed the  
occurrence of a superconducting ground state if the magnetic order is  
suppressed by external pressure ($p_c=2.8$~GPa)  
\cite{Grosche96,Mathur98,Raymond99,Demuer01,Shekin01,Raymond00a,Demuer02}.  
If the change of interatomic distances is the main source of altering  
the exchange coupling between $4f$ and conduction electrons, $J$, then  
CePd$_2$Ge$_2$ should reveal a pronounced variation of $T_N(p)$.  The  
interest in high pressure studies on CePd$_{2.02}$Ge$_{1.98}$ is to  
explore the role of stoichiometry on $J$ and thus on $p_c$. Measuring  
simultaneously $C(T)$ and $\rho(T)$ has the advantage that independent  
information about the strength of electronic correlations from the  
{\it same} specimen can be obtained.  
   
In order to draw a credible conclusion about the pressure response of  
both compounds, it is essential to expose both samples to the {\it  
same} pressure conditions. The best way to achieve this is to place  
both specimens adjacent to each other in the {\it same} pressure  
cell.   
   
%################################################################   
%   
%           EXPERIMENTAL DETAILS   
%   
%################################################################   
\section{\label{sec:experimental} Experimental Details}    
\subsection{Sample Preparation and Characterization}   
   
The CePd$_{2+x}$Ge$_{2-x}$ compounds have been prepared by melting Ce  
(4N), Pd (5N), and Ge (6N) according to the composition in an arc  
($x=0$) or an induction furnace ($x=0.02$) under Ar (6N)  
atmosphere. The samples have been melted several times to achieve good  
homogeneity. Mass loss during melting and annealing  
(CePd$_{2.02}$Ge$_{1.98}$ at 1420~K and CePd$_2$Ge$_2$ at 1470~K for  
two days) was negligible. A part of the polycrystalline ingots has  
been analyzed by x-ray powder diffraction. The diffraction pattern  
contained only peaks according to the ThCr$_2$Si$_2$ structure  
($I4/mmm$). The magnetic structure of CePd$_2$Ge$_2$ consists of  
ferromagnetic planes stacked antiferromagnetically along the  
[110]-direction with moments ($\mu=0.85\mu_{\rm B}$ at 1.8~K) parallel  
to the stacking direction \cite{Fak00,Dijk00,Feyerherm98}. No  
information about the magnetic structure of CePd$_{2.02}$Ge$_{1.98}$  
is available but it is very likely that the small Pd-excess does not  
change the antiferromagnetic structure.  
   
The measurements of the specific heat, dc-magnetic susceptibility, and  
electrical resistivity at ambient pressure revealed for both compounds  
a phase transition into an antiferromagnetically ordered phase at  
about 5.1~K (Tab.\ \ref{tableambientpressure}). The low temperature  
specific heat (0.3~K~$<T\leq$~2~K) can be described by the sum of an  
electronic ($\gamma T$) and a magnon-like $T^3$ part with a gap  
$\Delta$ in the excitation spectrum. The increased $\gamma$-value of  
the Pd-rich compound in respect to the stoichiometric one points to  
enhanced correlations. The $\gamma$ and $\Delta$ values of  
CePd$_2$Ge$_2$ obtained here are larger than those reported in Ref.\  
\cite{Besnus92} due to the enlarged temperature range accessible in  
the present study. However, the absolute values of $C_p$ at $T_{\rm  
N}$ are almost the same. The entropy release at $T_N$ is $S/R\approx  
0.8\ln 2$ and reaches $\ln 2$ at about 9~K for both compounds. The  
high temperature magnetic susceptibility can be described by a  
Curie-Weiss law $\chi\propto \mu_{\rm eff}/(T-\Theta)$, with an  
effective moment $\mu_{\rm eff}$ close to the free moment value and  
$\Theta$ the Curie-Weiss temperature.  The residual scattering is  
rather low in both compounds; the non-stoichiometric sample has the  
lower $\rho_0$-value.  
   
\begin{table}   
\caption{\label{tableambientpressure}Ambient pressure data of  
CePd$_{2.02}$Ge$_{1.98}$ and CePd$_2$Ge$_2$. The N\'eel temperature  
$T_N$ is the mean value of specific heat, dc-magnetic susceptibility,  
and electrical resistivity measurements. The specific heat can be  
described by $C=\gamma T +\beta T^3\exp(-\Delta/T)$ for  
$T\leq$~2~K. $\mu_{\rm eff}$ is the effective magnetic moment,  
$\Theta$ the Curie-Weiss temperature, $\rho_0$ the residual  
resistivity, and RRR the ratio $\rho(295~{\rm K})/\rho_0$.}  
 
\begin{ruledtabular}   
\begin{tabular}{lll}   
                      & CePd$_{2.02}$Ge$_{1.98}$  & CePd$_2$Ge$_2$ \\ \hline   
a (\AA)                  & 4.3399(7)              & 4.3411(5)         \\   
c (\AA)                  & 10.0343(19)            & 10.0417(5)        \\   
V (\AA$^3$)              & 189.00(7)              & 189.23(5)         \\    
$T_N$~(K)                & 5.16(8)                & 5.12(7)           \\    
$\gamma$(mJ/(mol~K$^2$)) & 101(5)                 & 44(1)             \\   
$\beta$(mJ/(mol~K$^4$))  & 148(5)                 & 234(6)            \\   
$\Delta$ (K)             & 0.8(1)                 & 1.6(1)            \\   
$\mu_{\rm eff}$($\mu_{\rm B}$) at 300 K & 2.1     & 2.5               \\   
$\Theta$ (K)             & -24(5)                 & -16(3)            \\               
$\rho_0$($\mu\Omega$cm)  & 1.4(1)                 & 1.7(1)            \\   
RRR                      & 32(3)                  & 29(3)             \\    
\end{tabular}                                  
\end{ruledtabular}   
\end{table}   
   
\subsection{High Pressure Set-Up}   
Samples with cross sections of $23\times 59~\mu$m$^2$ ($x=0$) and  
$22\times 48~\mu$m$^2$ ($x=0.02$) have been cut from the  
polycrystalline ingots and placed into the pressure chamber (internal  
diameter of 1~mm) \cite{Jaccard98}. A small piece of Pb served as  
pressure gauge \cite{Bireckhoven88} and was connected in series to the  
samples for four-point measurements. The samples have been arranged in  
the pressure chamber in such a way that the crystallographic $c$-axis  
was parallel to the pressurizing force. For the ac-calorimetry  
measurements the sample itself was used as heater and was thermally  
excited by an oscillating heating power $P=P_0[1+ \cos(\omega t)]$,  
due to an applied ac-voltage of frequency $\omega / 2$. At steady  
state it increases the sample temperature by $\Delta T$ above the bath  
temperature $T_0$. This temperature increase contains a  
time-independent offset $T_{\rm dc}=P_0/\Lambda$, with $\Lambda$ the  
thermal conduction of the heat link between sample and pressure cell  
(to a first approximation: the pressure-transmitting medium, i.e.,  
steatite). In ideal conditions the oscillatory part of $\Delta T$ is  
$T_{\rm ac}=P_0/(\omega C_p)$ \cite{Sullivan68}. These temperature  
oscillations have been measured with a \underline{Au}Fe/Au  
thermocouple attached to the sample. The thermovoltage $V_{\rm ac}$  
arises from the temperature difference between the sample (at $T_0 +  
\Delta T$) and the edge of the sample chamber (at $T_0$)  
\cite{Jaccard98}.  
   
The dynamic response of the sample involves two time constants  
$\tau_1=C_p/\Lambda$ and $\tau_2$. The former expresses the thermal  
coupling between the sample and the temperature bath, whereas the  
latter represents a characteristic time for the sample to reach  
thermal equilibrium. When the measuring frequency fulfills the  
condition $\omega \tau_1 > 1 \gg \omega \tau_2$, the ac-technique  
yields the specific heat of the sample. In the course of the  
experiment this condition was checked at several temperatures and  
pressures. The frequencies used were in the range 750~Hz $\leq \omega  
\leq 3000$~Hz. The condition $\omega \tau_2 \ll 1$ is  
fulfilled for metals because they ensure high thermal conductivity  
within the sample.  
   
The inverse of the recorded lock-in voltage $V_{\rm ac}$ is  
proportional to $C/T$, since the temperature dependence of the  
absolute thermoelectrical power, $S(T)\propto T$, is a fairly good  
assumption at $T< 1$~K. Above this temperature the $S(T)$ dependence  
is certainly different and $1/V_{\rm ac}$ has to be interpreted with  
caution. Nevertheless, the present set-up has several  
advantages. Firstly, it is possible to check whether a pronounced  
anomaly in $1/V_{\rm ac}$ is related to the sample or not with an  
independent $\rho(T)$ measurement on the {\it same} sample; secondly,  
it excludes an additional source of pressure inhomogeneity due to a  
heater attached to the sample; and thirdly, internal temperature  
gradients can be reduced as much as possible.  
   
With such an arrangement it is in principle possible to calibrate the  
\underline{Au}Fe/Au thermocouple up to very high pressure and over a  
wide temperature range \cite{Wilhelm02}. Here, we have only determined  
$S(T)$ at 4.2 K and 1.0 K to get a rough estimate of the influence of  
pressure on $S(T)$ \cite{au_treatment}. The obtained values of the  
absolute thermopower of \underline{Au}Fe at 4.2~K and 1.0~K at 12~GPa  
are about 20\% smaller than the values at ambient pressure. These  
rather small changes show that the interpretation of the results  
reported in this work is not affected qualitatively if the ambient  
pressure values of $S(T)$ are used.  
   
The sample chamber has been carefully re-examined after pressure  
release to rule out changes in the positions of the voltage  
leads connected to the samples. The overall shape of the pressure cell  
as well as its initial diameter were almost unchanged and the distance  
between the voltage leads increased by less than 5\%. Taking this  
uncertainty in the geometrical factor as well as the change in volume  
at high pressure into account the absolute value of $\rho(T)$ can be  
determined within 20\%.

%################################################################   
%   
%           RESULTS FROM TRANSPORT   
%   
%################################################################   
   
\section{\label{sec:results}Results}   
\subsection{\label{sec:transportresults}Transport measurements}   
\begin{figure}   
\includegraphics[width=86mm,clip]{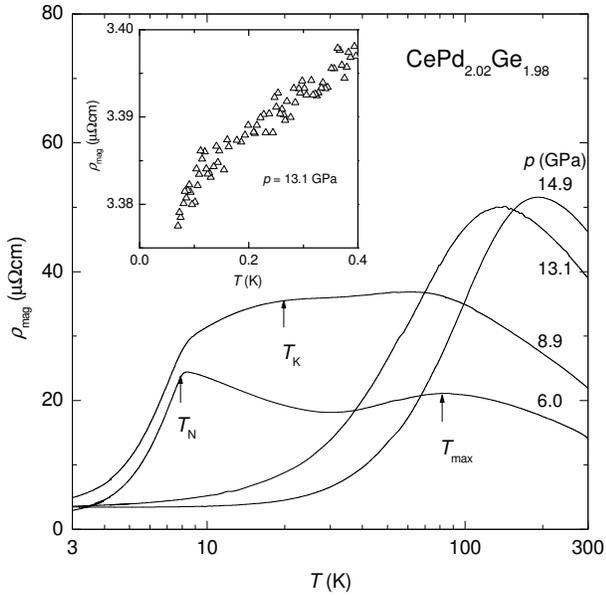}  
\caption{\label{rhovstcpxgx}The magnetic part $\rho_{\rm mag}(T)$ of  
the electrical resistivity of CePd$_{2.02}$Ge$_{1.98}$ vs temperature, $T$,  
in a semilogarithmic plot. The antiferromagnetic transition produces a  
cusp at $T_N$. The scattering of carriers at the ground state and  
excited crystal field levels produce two maxima at $T_K$ and $T_{\rm  
max}$. Inset: An additional phase transition at high pressure might be  
responsible for the decrease of $\rho_{\rm mag}(T)$ at about 110~mK.}  
\end{figure}

\begin{figure}   
\includegraphics[width=86mm,clip]{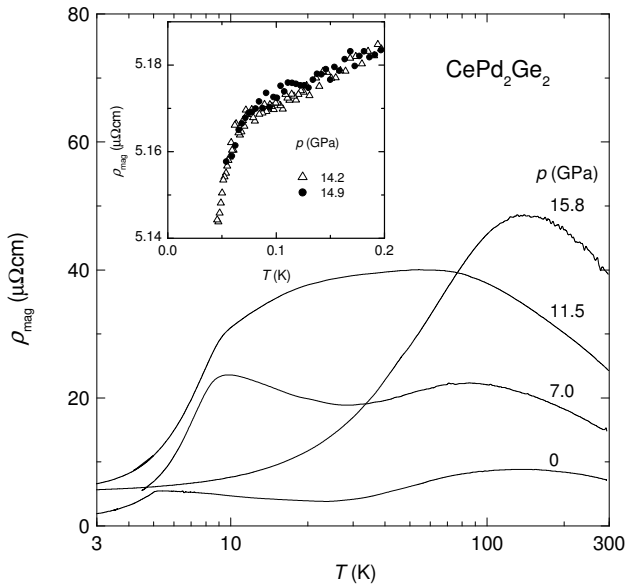}  
\caption{\label{rhovstcpg}Temperature dependence of the magnetic part  
of the electrical resistivity $\rho_{\rm mag}(T)$ of CePd$_2$Ge$_2$ in  
a semilogarithmic plot. The similarity of the high pressure curves to  
those of CePd$_{2.02}$Ge$_{1.98}$ (cf. Fig.\ \ref{rhovstcpxgx}) is  
evident. The small drop in $\rho_{\rm mag}(T)$ at 70~mK above 14~GPa  
(inset) might indicate an additional phase transition.}  
\end{figure}

The entrance into the magnetically ordered state is clearly visible by 
a cusp in $\rho_{\rm mag}(T)$ of CePd$_{2.02}$Ge$_{1.98}$ and 
CePd$_2$Ge$_2$ (Fig.\ \ref{rhovstcpxgx} and Fig.\ 
\ref{rhovstcpg}). The magnetic contribution $\rho_{\rm mag}(T)$ to 
$\rho(T)$ has been obtained by subtracting a phononic contribution, 
approximated as $\rho_{\rm ph}(T)=0.1~\mu\Omega{\rm cm/K}\times T$, 
from the raw data.  Qualitatively, both compounds show the same 
pressure dependence: Pressure shifts $T_N$ upwards, the signature of 
the phase transition broadens, and within a small pressure range, the 
traces of the phase transition vanish. $\rho_{\rm mag}(T)$ exhibits 
two maxima, reflecting the Kondo scattering from the ground state and 
excited crystal field levels as often observed for other compounds 
(indicated by $T_K$ and $T_{\rm max}$, respectively, in Fig.\ 
\ref{rhovstcpxgx}) \cite{Wilhelm00,Jaccard98}.  Furthermore, a small 
and reproducible decrease in $\rho_{\rm mag}(T)$ has been detected at 
very low temperature (insets of Fig.\ \ref{rhovstcpxgx} and Fig.\ 
\ref{rhovstcpg}). It occurs in a narrow pressure range in the 
apparently non-magnetic phase below 110~mK for 
CePd$_{2.02}$Ge$_{1.98}$ whereas it was found in CePd$_2$Ge$_2$ at 
somewhat lower temperature (70~mK). In each case an increased 
measuring current density suppressed the anomaly. 
   
\begin{figure}   
\includegraphics[width=86mm,clip]{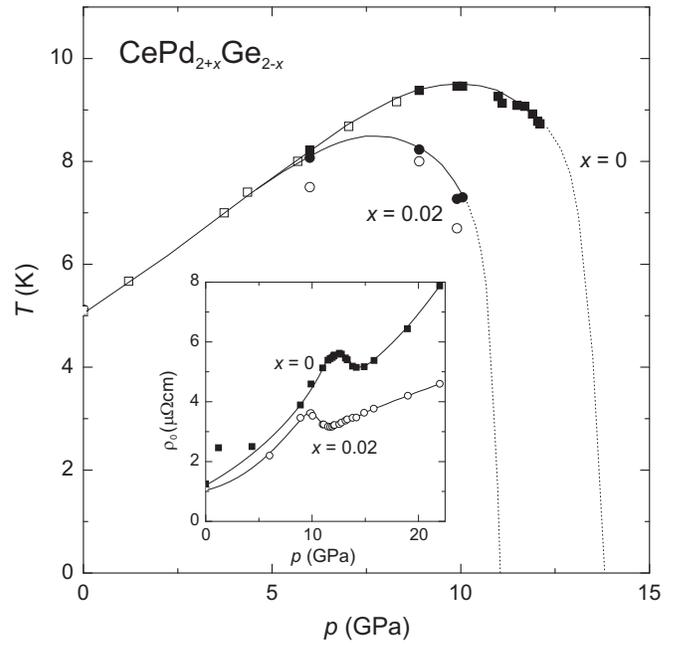}   
\caption{\label{tnvsp}Pressure dependence of $T_N$ for  
CePd$_{2+x}$Ge$_{2-x}$ ($x=0$ and 0.02). The extrapolation  
$T_N\rightarrow 0$ assumes critical pressures $p_c=11.0$~GPa  
($x=0.02$) and 13.8~GPa ($x=0$). Two data sets of $T_N(p)$ for $x=0$  
obtained on different samples match perfectly. The $T_N$ values for  
$x=0.02$ obtained by the ac-calorimetry are represented by open  
circles. Inset: A pronounced variation of the residual resistivity  
$\rho_0$ with pressure is observed in both compounds.}  
\end{figure}

These measurements reveal the pressure dependence of $T_N$, as depicted  
in Fig.\ \ref{tnvsp}. As criterion for $T_N$ the intersection of two  
tangents drawn to $\rho(T)$ has been used. In the case of  
CePd$_2$Ge$_2$ both data sets obtained on different samples from the  
same batch in different pressure cells match perfectly (open and  
filled squares in Fig.\ \ref{tnvsp}). The initial pressure shift  
$\partial T_N/\partial p= 0.51(1)$~K/GPa is slightly higher than that  
reported in Ref.\ \cite{Oomi96}. The same value is obtained for  
CePd$_{2.02}$Ge$_{1.98}$ if the values of $T_N$ at ambient pressure  
and 6~GPa are used. At higher pressures however, both  
$T_N(p)$-variations are clearly different. In  
CePd$_{2.02}$Ge$_{1.98}$, $T_N$ does not reach the same absolute value  
as in the stoichiometric compound and the magnetic order vanishes  
already at $p_c=11.0$~GPa, compared to $p_c=13.8$ GPa for  
CePd$_2$Ge$_2$. Thus, the Pd-excess has led to a reduction of $p_c$  
by 2.8~GPa.  
%################################################################   
%   
%           FIGURE: RHO(T) OF CePd2.02Ge1.98   
%   
%################################################################   
%################################################################   
%   
%           FIGURE: RHO(T) OF CePd2Ge2   
%   
%################################################################   
%################################################################   
%   
%           FIGURE: T_N vs PRESSURE   
%   
%################################################################   
%################################################################   
%   
%          FIGURE: A VS P, INSET N VS P FOR BOTH SAMPLES   
%   
%################################################################   
\begin{figure}   
\includegraphics[width=86mm,clip]{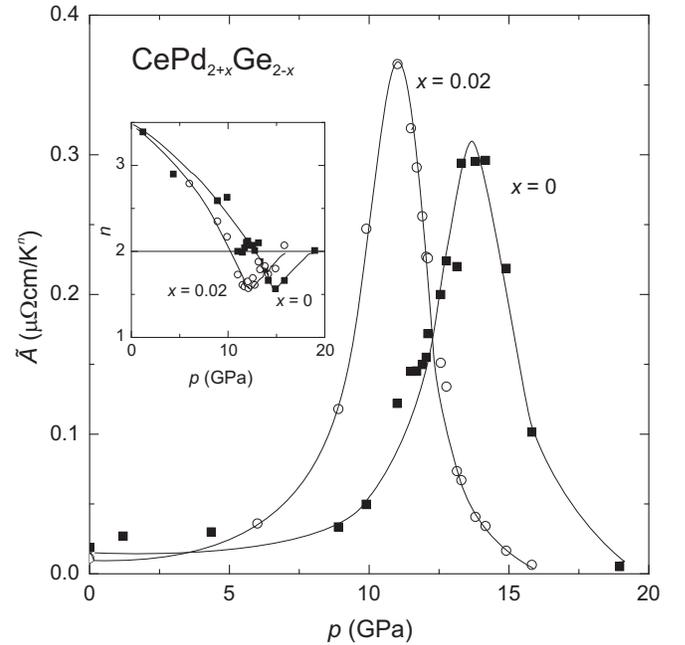}   
\caption{\label{avsp}Temperature coefficient $\tilde{A}$ obtained from  
a fit $\rho_{\rm mag}(T)=\rho_0 + \tilde{A}T^n$ to the data of  
CePd$_{2+x}$Ge$_{2-x}$ ($x=0$ and 0.02) below 2~K vs pressure, $p$. $\tilde{A}$ peaks  
at 11.0~GPa and 13.8~GPa for $x=0.02$ and $x=0$, respectively. Inset:  
The exponent $n$ used to describe $\rho_{\rm mag}(T)$ at different  
pressures.}  
\end{figure}   
   
The extrapolation $T_N \rightarrow 0$ (dotted lines in Fig.\  
\ref{tnvsp}) is based on the assumption that the maximum in the  
$\tilde{A}(p)$ dependence can be taken as critical pressure. The  
$\tilde{A}$-values have been obtained from fits of $\rho_{\rm  
mag}(T)=\rho_0 +\tilde{A} T^n$ to the data below 2~K, with $\tilde{A}$  
and $n$ as fitting parameters. The lower bound to the fit is given by  
the accessible temperature and is about 40~mK. The value for the upper  
limit is a compromise between an as narrow as possible temperature  
interval and the reliability of the deduced parameters. The pressure  
dependence of the temperature coefficient $\tilde{A}$ is qualitatively  
similar for both compounds and shows a pronounced anomaly which is  
assumed to be the hallmark of the magnetic/non-magnetic phase  
transition (see Fig.\ \ref{avsp}). Both $\tilde{A}(p)$-variations can be  
mapped on top of each other if a pressure shift of $\Delta p_{{\rm  
max}\{\tilde{A}\}}=2.8$~GPa is taken into account. In the magnetically  
ordered phase the exponent $n>2$ whereas $n=2$ is  
found in the Fermi-liquid regime far above $p_c$ (inset Fig.\  
\ref{avsp}). Exponents smaller than two are observed in a certain  
pressure range around $p_c$. A minimum $n\approx 1.6$ is attained just  
above $p_c$ for both compounds. Within a small pressure range around $p_c$,  
$\rho_{\rm mag}(T)$ cannot be  
described by a quadratic temperature dependence even if the  
temperature interval is 40~mK$<T<0.6$~K. Similar observations have  
been reported for CePd$_2$Si$_2$ \cite{Demuer01,Demuer02} and other  
systems, like CeRu$_2$Ge$_2$ \cite{Wilhelm99} and CeCu$_5$Au  
\cite{Wilhelm00}.  
   
Approaching the verge of magnetism seems also to affect the residual  
resistivity $\rho_0$. It is very sensitive to small pressure changes  
and exhibits anomalies around $p_c$ which are qualitatively the same  
for both compounds (see inset Fig.\ \ref{tnvsp}). Just below $p_c$,  
$\rho_0$ attains a local maximum and passes through a local minimum  
above $p_c$. Upon further pressure increase $\rho_0$ continuously  
increases and at 22~GPa reaches several times its ambient pressure  
value. This variation reflects intrinsic effects since a change of the  
geometrical factor in such a peculiar manner can be ruled out.  
  
\begin{table}   
\caption{\label{tablehighpressure}Pressure values where the residual  
resistivity $\rho_0$ and the fitting parameter of $\rho_{\rm mag}(T)$  
as well as $1/V_{\rm ac}$ for $T\rightarrow 0$ show anomalies. At  
$p_c^{\rm cal}$ the calculated unit cell volume of CePd$_2$Ge$_2$ is  
equal to that of CePd$_2$Si$_2$ at its critical pressure  
($p_c=3.9$~GPa \cite{Demuer02}). Each anomaly occurring in  
CePd$_{2.02}$Ge$_{1.98}$ is also seen in CePd$_2$Ge$_2$ but shifted by  
$\Delta p$ as indicated in the last column.}  
\begin{ruledtabular}   
\begin{tabular}{lcc|c}   
                              &CePd$_{2.02}$Ge$_{1.98}$ & CePd$_2$Ge$_2$ & $\Delta$ $p$(GPa) \\ \hline   
$p_{\rm{max}\{\rho_0\}}$ (GPa)& 9.7            & 12.3                 & 2.6               \\   
$p_c=p_{{\rm{max}}\{\tilde{A}\}}$ (GPa)& 11.0    & 13.8                 & 2.8               \\   
$p_{{\rm{min}}\{\rho_0\}}$ (GPa)& 11.6           & 14.5                 & 2.9               \\   
$p_{{\rm{max}}\{(1/V_{\rm ac})_{T\rightarrow 0}\}}$ (GPa) &11.7&--          & --                \\   
$p_{{\rm{min}}\{n\}}$ (GPa)   & 12.1             & 14.6                 & 2.5               \\   
$p_c^{\rm cal}$ (GPa)           & 14.2             & 14.4                 & 0.2              \\   
\end{tabular}                                  
\end{ruledtabular}   
\end{table}   
   
The comparison of the resistivity data presented above reveals that  
pressure has qualitatively the same effect on both compounds. The main  
difference is that for CePd$_{2.02}$Ge$_{1.98}$ less pressure ($\Delta  
p=2.7(2)$~GPa) is necessary to achieve the same effect as in  
CePd$_2$Ge$_2$.  Table~\ref{tablehighpressure} summarizes several  
quantities which show pronounced anomalies in their pressure  
behavior. From this we infer that not only interatomic distances are  
important for $J$ since the difference in unit cell volume at ambient  
pressure cannot explain such a large shift in $p_c$.  
   
%################################################################   
%   
%           RESULTS FROM SPECIFIC HEAT   
%   
%################################################################   
\subsection{\label{sec:accalorimetry}\lowercase{ac}-calorimetry on CePd$_{2.02}$Ge$_{1.98}$ }   
\begin{figure}   
\includegraphics[width=86mm,clip]{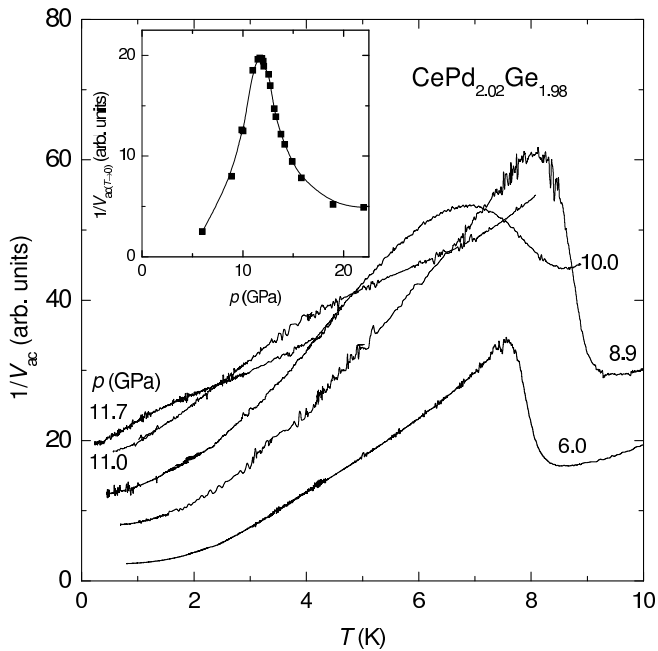}    
\caption{\label{cpxgxlowp}Temperature dependence of the inverse  
lock-in voltage $V_{\rm ac}$ of CePd$_{2.02}$Ge$_{1.98}$. The entrance  
into the antiferromagnetically ordered state is clearly  
visible. Inset: The values of 1/$V_{\rm ac}\propto C/T$ taken at low  
temperature show a pronounced peak in the vicinity of $p_c$.}  
\end{figure}

Figure~\ref{cpxgxlowp} shows the inverse of the registered lock-in  
signal $V_{\rm ac}$ below 10 K at various pressures. The pronounced  
anomaly in $1/V_{\rm ac}(T)$ for pressures between 6.0 GPa and 10 GPa  
is caused by the entrance into the antiferromagnetically ordered  
phase. Taking the temperature of the maximum as $T_N$ yields lower  
$T_N$-values as those shown in Fig.\ \ref{tnvsp}. $T_{\rm N}$ taken  
from the midpoint of the $1/V_{\rm ac}$-anomaly at 6~GPa is the same as that  
deduced from $\rho(T)$. At higher pressure however, this definition  
yields larger $T_N$-values than those obtained from $\rho(T)$. The  
height of the anomaly decreases and it becomes a very broad feature as  
the system approaches $p_c$. A similar broadening upon approaching  
$p_c$ has been reported for CeRu$_2$Ge$_2$ \cite{Demuer00} and  
CePd$_2$Si$_2$ \cite{Demuer01} examined in pressure cells with solid  
He as pressure transmitting medium, despite their lower  
$p_c$-values.  
   
From a general point of view this might be due to inhomogeneous  
pressure conditions always present regardless the pressure medium and  
the absolute value of $p_c$. Close to $p_c$ the $T_N(p)$ variation is  
very strong and a small pressure gradient can easily generate $\Delta  
T_N\approx 1$~K. However, other intrinsic effects cannot be excluded  
to be responsible for a broadening in the vicinity of $p_c$.  
   
A very interesting observation is the pressure dependence of the value  
of $1/V_{\rm ac}$ taken at the lowest temperature reached in each  
pressure run (inset Fig.\ \ref{cpxgxlowp}). Upon approaching $p_c$ it  
strongly increases, reaches a maximum just above $p_c$ (which was  
inferred from the $\tilde{A}(p)$-anomaly), and levels off at high  
pressure. As was pointed out above, $1/V_{\rm ac}(T)\propto C/T$ at  
low temperature can be regarded as a direct measure of the electronic  
correlations. The pronounced pressure dependence of $1/V_{\rm ac}$  
shows that the electronic correlations are considerably enhanced as  
pressure approaches $p_c$ and that the signal originates mainly from  
the sample.  
%################################################################   
%   
%          FIGURE: C/T OF CePdxGex   
%   
%################################################################   
   
%################################################################   
%   
%                   DISCUSSION   
%   
%################################################################   
\section{\label{sec:discussion}Discussion}    
In the following we will first compare the $T_N(V)$-dependence of  
CePd$_2$Ge$_2$ with that of CePd$_2$Si$_2$. Thereafter, the elaborate  
discussion of the pressure effects on the CePd$_{2+x}$Ge$_{2-x}$  
compounds ($x=0$ and 0.02) will reveal a possible explanation of the  
observed similarities as well as the differences.  
   
The $(T,p)$ phase diagram of each CePd$_{2+x}$Ge$_{2-x}$ compound 
($x=0$ and 0.02) presented in Fig.\ \ref{tnvsp} can be qualitatively 
understood within the Doniach picture \cite{Doniach77}. Pressure tunes 
the characteristic energy scales $T_K \propto 
\exp(-1/Jn(E_F))$ and $T_{\rm RKKY} \propto (Jn(E_F))^2$, involved in 
the Kondo effect and the RKKY interaction, respectively. Here $J$ is the 
exchange coupling between 4$f$ and conduction electrons and $n(E_F)$ 
is the density of states at the Fermi Energy $E_F$. The RKKY 
interaction dominates the Kondo effect for small $Jn(E_F)$-values as 
in CePd$_2$Ge$_2$ at low pressures where $T_K$ is very small.  The 
slight difference in composition has little effect on $T_N$ at ambient 
pressure and its pressure dependence below 6~GPa. In both samples 
$Jn(E_{\rm F})$ is enhanced by pressure and forces the system into a 
non-magnetic state for $p>p_c$. 
   
\begin{figure}   
\includegraphics[width=86mm,clip]{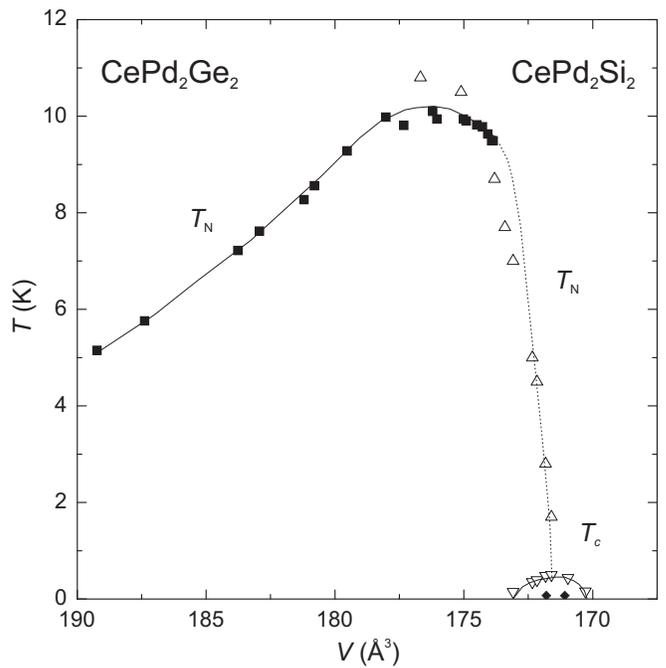}    
\caption{\label{tnvsv}Transition temperatures of CePd$_2$Ge$_2$ and  
CePd$_2$Si$_2$ as a function of a calculated unit cell volume, $V$. $T_N$  
and $T_c$ represent the N\'eel temperature and the superconducting  
transition temperature, respectively. The data of CePd$_2$Si$_2$ are  
taken from Ref.\ \cite{Demuer02}. As $T_N$-criterion the temperature  
derivative of $\rho_{\rm mag}(T)$ was used.}  
\end{figure}

If the unit cell volume is the crucial parameter which determines  
$T_N$ then it should be possible to plot the $T_N(p)$-data of  
CePd$_2$Ge$_2$ and its Si-counterpart CePd$_2$Si$_2$ in a common  
$(T,V)$ phase diagram. As pressure has tuned the unit cell volume of  
CePd$_2$Ge$_2$ to that of CePd$_2$Si$_2$ both $T_N$-values should be  
comparable as the $T_{\rm N}(V)$-variation of the solid-solution  
CePd$_2$(Si$_{1-x}$Ge$_x$)$_2$ suggest \cite{Das91}. The  
transformation of pressure into volume was done with a bulk modulus   
$B_0=120$~GPa (and $B'=4$ for its pressure dependence)  
for both compounds. This is a reasonable assumption for ternary Ce-compounds,  
crystallizing in the ThCr$_2$Si$_2$-type of structure as was pointed out  
in Ref.\ \cite{Wilhelm99}. We used as a unit cell volume of  
CePd$_2$Si$_2$ at ambient pressure $V_0=176.83$~\AA$^3$ which is   
the mean value of the literature data  
\cite{Raymond99,Steeman88,Dijk00,Rossi79,Link96}. Figure \ref{tnvsv}  
exhibits a pronounced $T_N(V)$-variation, starting at $T_{\rm  
N}=5.1$~K, passing through a maximum of about 10~K, and eventually  
approaching zero. The extrapolation $T_N\rightarrow 0$ yields  
$V=171.51~$\AA$^3$ which correspond to a pressure of 14.4~GPa, very  
close to $p_c=13.8$~GPa deduced from the maximum in  
$\tilde{A}(p)$. CePd$_2$Si$_2$ reaches this volume at  
$p_c=3.9$~GPa. This $p_c$-value agrees perfectly with that found in a  
thorough investigation of the strain enhancement of superconductivity  
\cite{Demuer02}.  
   
%################################################################   
%   
%          FIGURE: T_N of CPG and CPS vs Volume   
%   
%################################################################   
   
The almost perfect match of the two data sets emphasizes the  
importance of the sample orientation with respect to the direction of  
the applied force. Figure~\ref{tnvsv} shows only data from samples with  
their crystallographic $c$-axis parallel to the applied force exerted  
to the anvils. The additional uniaxial strain component along the  
$c$-axis strongly affects the anisotropy of the tetragonal  
ThCr$_2$Si$_2$ structure. It is known from x-ray absorption studies  
\cite{Sampathkumaran85} that a change of the $c$-axis lattice  
parameter results in a varying chemical bonding strength between the  
Ce-ions and their ligands. A change in the Ce-valence, however, is  
thought to influence only the $a$-axis lattice parameter and vice  
versa. Thus, it seems that the additional uniaxial stress is necessary  
to change the hybridization in an effective way. In CePd$_2$Si$_2$ it  
shifts $p_c$ from 2.8~GPa to 3.9~GPa and leads to an increase in the  
superconducting transition temperature of about 40\% \cite{Demuer02}  
compared to values obtained from samples in a configuration where the  
crystallographic $c$-axis was perpendicular to the external force  
\cite{Grosche96,Mathur98,Demuer01,Shekin01}.  
   
The combined phase diagram clearly demonstrates the analogy between  
CePd$_2$Ge$_2$ at high pressure and CePd$_2$Si$_2$ at moderate  
pressures. It can provide an idea about the sudden decrease of  
$\rho_{\rm mag}(T)$ in CePd$_{2.02}$Ge$_{1.98}$ and CePd$_2$Ge$_2$ at  
110~mK and 70~mK, respectively (insets Fig.\ \ref{rhovstcpxgx} and  
Fig.\ \ref{rhovstcpg}). The reduced volume where these anomalies occur  
is similar to the volume where superconductivity is found in  
CePd$_2$Si$_2$ (Fig.\ \ref{tnvsv}). An interpretation as incipient  
superconductivity is therefore one possible explanation, especially if  
the reported properties of CeNi$_{2+x}$Ge$_{2-x}$  
\cite{Steglich00,Jaccard01,Grosche00} and CePd$_2$Si$_2$  
\cite{Demuer01,Grosche01} are recalled. In the former system $\rho=0$  
was only achieved after $\rho_0$ had been reduced below  
1-2~$\mu\Omega$cm. For those samples with higher $\rho_0$-values only  
traces of superconductivity appeared. Also for the latter compound  
high-purity samples ($\rho_0<1\mu\Omega$cm) seem to be required for a  
complete superconducting transition \cite{Mathur98}.  
   
The difference in the $p_c$-values and the $\tilde{A}(p)$-dependence  
of CePd$_{2.02}$Ge$_{1.98}$ and CePd$_2$Ge$_2$ can also be understood  
qualitatively within the Doniach picture. Magnetic order is a  
cooperative phenomenon involving the alignment of spins over distances  
which are large compared to the lattice parameter. It is very unlikely  
that a small change in the Ce-ligand configuration will influence the  
RKKY interaction and $J \approx J'$ seems to be justified, with $J$  
and $J'$ the exchange coupling for $x=0$ and $x=0.02$,  
respectively. The almost identical $T_N$-values at ambient pressure  
corroborate this assumption. For the Kondo effect however, only the  
local environment of the Ce-ions is essential and therefore, a small  
change in its configuration sphere should influence $T_K$. It seems  
very likely that the additional Pd in CePd$_{2.02}$Ge$_{1.98}$  
occupies Ge-sites and/or interstitial sites and might influence the  
local environment of the Ce-ions, resulting in $J'>J$. Within a  
certain limit this hypothesis is supported by the different  
$\gamma$-values at ambient pressure (Tab.\  
\ref{tableambientpressure}).  Furthermore, the x-ray data of  
CePd$_{2.02}$Ge$_{1.98}$ reveal that the $c$-axis lattice parameter is  
slightly lower, whereas the $a$-axis lattice parameter is unchanged  
within the standard deviation. Thus, the Pd-excess could have caused a  
stronger hybridization between the $4f$ and conduction electrons  
already at ambient pressure as was argued above. Results of a thorough  
investigation of the influence of the Ni-excess in  
Ce$_{1.005}$Ni$_{2+z}$Ge$_{2-z}$ encourage this argumentation  
\cite{Steglich00}. As a function of the Ni-content the $c$-axis  
lattice parameter and $\rho_0$ showed a minimum at $z=0.02$ and a  
complete superconducting transition occurred close  
to this value at ambient pressure, whereas pressure had to be applied  
to achieve a superconducting ground state in the stoichiometric  
compound \cite{Braithwaite00}. With these assumptions $T_{\rm  
RKKY}(J)$ will be of comparable strength to $T_K(J)$ and $T_{\rm  
K}(J')$ at a critical value $J'_c<J_c$. Therefore, $p_c$ of  
CePd$_{2.02}$Ge$_{1.98}$, determined by $J'_c$, is lower than $p_c$ of  
CePd$_2$Ge$_2$, given by $J_c$. As a consequence of this consideration  
a shifted $\tilde{A}(p)$-dependence follows, since $\tilde{A}\propto  
T_K^{-2}$, as is experimentally found and depicted in Fig.\  
\ref{avsp}.  
%################################################################   
%   
%          FIGURE: T_K high and low vs p   
%   
%################################################################   
\begin{figure}   
\includegraphics[width=86mm,clip]{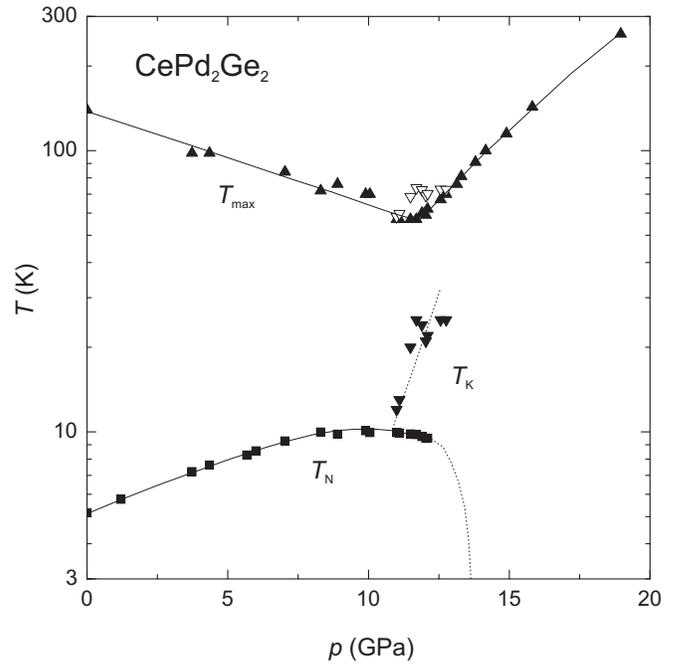}    
\caption{\label{phasediagram}Pressure dependence of some  
characteristic temperatures, $T_{\rm max}$, $T_N$, and $T_K$ of  
CePd$_2$Ge$_2$ in a semilogarithmic plot. The pressure dependence of  
$T_{\rm max}$ shows a minimum close to $p_c$. The open symbols  
represent calculated data of $T_K^h$ (see text).}  
\end{figure}    
   
The Kondo temperature in CePd$_2$Ge$_2$ at low pressures is small in  
comparison to the crystal field splitting $\Delta_1 =110$~K and  
$\Delta_2 = 220$~K \cite{Besnus92}. Therefore, $\rho_{\rm mag}(T)$  
shows only one maximum at $T_{\rm max}$ at low pressure. Around $p_c$  
however, two maxima at $T_K$ and $T_{\rm max}$ occur, reflecting the  
Kondo scattering from the ground state and excited state, respectively  
(Fig.\ \ref{rhovstcpxgx} and Fig.\ \ref{rhovstcpg}). The low  
temperature maximum emerges at a pressure of about 11~GPa and shifts  
to higher temperatures with increasing pressure as shown in Fig.\  
\ref{phasediagram}. This might reflect an enhanced screening of the  
magnetic moments by the conduction electrons and thus point to an  
increasing role of the Kondo effect. As a consequence, the anomaly at  
low temperature has to be related to $T_K$. Both anomalies in  
$\rho_{\rm mag}(T)$ seem to merge in the vicinity of $p_c$, indicating  
the entrance into the intermediate valence regime. In this region the  
crystalline electrical field levels cease to exist as well-defined  
excitations. In the case of two excited crystal field levels, Hanazawa  
et al. \cite{Hanazawa85} have introduced a second Kondo temperature  
$T_K^h$ at high temperature. It is related to $T_K$ by  
$T_K^h=\sqrt[3]{T_K\Delta_1\Delta_2}$. Using the assumption that  
$\Delta_1$ and $\Delta_2$ decrease under pressure with the same rate  
as in CePd$_2$Si$_2$ \cite{Dijk00} the $T_K^h$-values can be  
calculated as a function of pressure (open symbols in Fig.\  
\ref{phasediagram}). A good agreement is achieved with the measured  
values of $T_{\rm max}$. Similar considerations can be made for  
CePd$_{2.02}$Ge$_{1.98}$, resulting in the same phase diagram but  
shifted by 2.8~GPa to lower pressures.  
   
The analysis of the low temperature behavior of $\rho_{\rm mag}(T)$  
have revealed almost the same anomalies in $\rho_0$, $\tilde{A}$, and  
$n$ for CePd$_{2.02}$Ge$_{1.98}$ and CePd$_2$Ge$_2$. Of particular  
interest is the strong pressure dependence of $\rho_0$ (inset Fig.\  
\ref{tnvsp}). It is an additional example of a pressure dependent  
residual resistivity scattering in HF compounds already pointed out in  
Ref.\ \cite{Jaccard99}. The anomalies in $\rho_0(p)$ cannot be caused  
by lattice defects and impurities alone like in conventional metals.  
The independent-electron approximation ($\rho_0 \propto 1/(k_F^2 l)$,  
with $k_F$ the Fermi-wavenumber and $l$ the mean free path) suggests  
that pressure should affect $\rho_0$ only weakly since both $k_F$ and  
$l$ react upon pressure only through the small change of electron  
density and interatomic distances. In metals with strongly interacting  
electrons and magnetic order the contributions to $\rho_0$ are not  
well understood. So far, a noticeable $\rho_0(p)$-dependence has been  
found in several HF systems either in the magnetic phase (CeCu$_5$Au  
\cite{Wilhelm01} and YbCu$_2$Si$_2$ \cite{Alami98}), close to $p_c$  
(CeAl$_3$ \cite{Jaccard95}) or in the para\-magnetic phase  
(CeCu$_2$Ge$_2$ and CeCu$_2$Si$_2$ \cite{Jaccard99}). Following the  
suggestion by Miyake and Maebashi \cite{Miyake01} quantum critical  
fluctuations should give rise to an enhanced impurity potential. It  
leads to an increase of $\rho_0$ through non-magnetic impurity  
scattering near a ferromagnetic or antiferromagnetic quantum critical  
point if many-body corrections of scattering are taken into  
account. This possibility and the wealth of $\rho_0(p)$ anomalies  
reported so far might indicate that only a part of $\rho_0$ is due to  
static disorder and that the large variation of $\rho_0(p)$ is an  
intrinsic property of a weakly disordered Kondo lattice  
\cite{Wilhelm01}.  
   
The deviation from a FL behavior around $p_c$ is an established fact  
and can be understood in the framework of spin fluctuation theory  
\cite{Millis93,Moriya95}. For spin fluctuations with 3D character  
$\rho(T)=\rho_0 +\tilde{A}T^n$, with $n=1.5$, is expected. The minimum  
values of $n$ depicted in Fig.\ \ref{avsp} are close to this value. It  
is considerably different from a linear temperature dependence which  
is expected for a distribution of Kondo temperatures  
\cite{Bernal95}. Therefore, Kondo disorder seems to be negligible.  
   
The ac-calorimetry data revealed a pronounced variation of the  
ac-signal recorded below 1~K. It was argued above that $1/V_{\rm ac}$  
can be taken as the linear coefficient of the specific heat,  
$C/T=\gamma$.  Its pressure dependence is not strong enough to follow  
the $\tilde{A}(p)$-dependence according to the empirical  
Kadowaki-Woods relation \cite{Kadowaki86}. Especially at pressures  
above 15~GPa, the low temperature value of $1/V_{\rm ac}$ decreases  
less than expected from $\tilde{A}(p)$ \cite{Wilhelm02}. A possible  
reason for this deviation might be the unknown thermal properties of  
the pressure transmitting medium (and perhaps also of the sample) at  
high pressure. They have changed significantly, which was only  
accounted for by adjusting the measuring frequency. A step towards a  
quantitative measure of $C_p$ at these conditions would be to achieve  
a control of the supplied heating power and the thermal contact  
between sample and pressure transmitting medium. Nevertheless, the  
strong pressure dependence of $1/V_{\rm ac}$ at low temperature is  
reminiscent to $\tilde{A}(p)$ and is a motivation for further studies.  

\newpage

%##############################################################   
%   
%                               CONCLUSION   
%   
%##############################################################   

\section{Conclusion}   
We reported results of a combined electrical resistivity, $\rho(T)$,  
and ac-calorimetry, $C(T)$, investigation of the antiferromagnetically ordered  
CePd$_{2.02}$Ge$_{1.98}$ ($T_N=5.16$~K) and $\rho(T)$ measurements of  
CePd$_2$Ge$_2$ ($T_N=5.12$~K) for pressures up to  
22~GPa. Both measuring techniques have been assembled in {\it one}  
Bridgman-type of high pressure cell. The particular sample arrangement  
guarantees similar pressure conditions essential for a comparison of  
the pressure induced effects. The ac-calorimetry and $\rho(T)$-data  
have been obtained from the {\it same} sample which is important to  
demonstrate the feasibility of the ac-technique at these extreme  
conditions. Both methods reveal a suppression of magnetic order at a  
critical pressure $p_c=11.0$~GPa and $p_c=13.8$~GPa for  
CePd$_{2.02}$Ge$_{1.98}$ and CePd$_2$Ge$_2$, respectively. The inverse  
of the ac-signal $1/V_{\rm ac}\propto C/T$ recorded at the lowest  
temperature exhibits an anomaly in the vicinity of $p_c$, reminiscent  
to $\tilde{A}(p)$, the temperature coefficient of $\rho(T)$. Although  
the pressure dependence of $1/V_{\rm ac}$ is not strong enough to  
follow the entire $\tilde{A}(p)$ dependence according to the  
Kadowaki-Woods relation it is evident that the ac-signal mainly  
represents the sample properties. These observations demonstrate the  
sensitivity of the ac-calorimetry to electronic correlations despite  
the small sample mass (some $\mu$g). From the combined $(T,V)$ phase  
diagram of CePd$_2$Ge$_2$ and CePd$_2$Si$_2$ it was concluded that  
interatomic distances play a crucial role for the hybridization  
between the $4f$ and conduction electrons, i.e., for the exchange  
coupling $J$, in the stoichiometric compound. In order to explain the  
large difference in the $p_c$-values it was argued that the  
Ce-coordination sphere in CePd$_{2.02}$Ge$_{1.98}$ has changed due to  
the Pd-excess in respect to CePd$_2$Ge$_2$. This affects the Kondo  
temperature, $T_K$, and therefore $T_K$ will be comparable to the RKKY  
interaction at a lower critical value of $J$ for  
CePd$_{2.02}$Ge$_{1.98}$. With this assumption the shifted  
$\tilde{A}(p)$-variations have also been explained. The deviation of  
$\rho(T)$ from a Fermi-liquid behavior in the vicinity of $p_c$ can be  
ascribed to 3D spin fluctuations. The strong variation of the residual  
resistivity $\rho_0$ with pressure around $p_c$ might indicate that  
only a part of $\rho_0$ is due to static disorder. As a consequence,  
the assumption of a powerlaw for $\rho(T)$ will be a subject of  
further careful investigations, especially at very low temperatures.  
   
%####################################################################################   
%   
%                               ACKNOWLEDGEMENT   
%   
%####################################################################################   
\begin{acknowledgments}   
We thank Y. Wang and A. Junod for fruitful discussions about the  
ac-calorimetric technique and R. Cartoni for technical support. The  
assistance of N. Caroca-Canales, H. Rave, and Z. Hossain at the MPI  
CPfS is acknowledged. We are grateful to A. Demuer for forwarding us  
the data of CePd$_2$Si$_2$ prior publication.  
\end{acknowledgments}   
  
\bibliography{heavyfermion}   

\begin{thebibliography}{52}
\bibitem[{au_()}]{au_treatment}
\bibinfo{note}{We have neglected the absolute thermopower of the pure Au--wire
  which has been annealed (3h at 1073 K) in air in order to oxydize
  uncontrolled Fe--impurities.}
\end{thebibliography}
 
\end{document}